\documentclass[prb,amsmath,amssymb,superscriptaddress,twocolumn]{revtex4}
\usepackage{float}
\usepackage{amsmath}
\usepackage{amssymb}
\usepackage{amsfonts}
\usepackage{euscript}
\usepackage{enumerate}
\usepackage{hhline}
\usepackage{pslatex}
\usepackage{tabularx}
\usepackage[usenames,dvipsnames]{xcolor}

\usepackage{graphicx}

\usepackage{dcolumn}
\usepackage{bm}
\usepackage[sort&compress]{natbib}

\usepackage{lipsum}

\makeatletter
\renewcommand{\@biblabel}[1]{#1. }
\renewcommand{\@dotsep}{500}
\renewcommand{\@pnumwidth}{0em}
\renewcommand{\l@figure}[2]{
\@dottedtocline{1}{1.5em}{2em}{Figure #1}{}\vspace{15pt}}

\newcommand{\ks}[1]{\textcolor{black}{#1}}
\newcommand{\xl}[1]{\textcolor{black}{#1}}
\newcommand{\xlt}[1]{\textcolor{black}{#1}}
\newcommand{\xlpr}[1]{\textcolor{black}{#1}}
\newcommand{\ys}[1]{\textcolor{black}{#1}}
\newcommand{\yst}[1]{\textcolor{black}{#1}}

\usepackage[normalem]{ulem}
\usepackage{upgreek}

\begin{document}
\title{Multi-mode microcavity frequency engineering through a shifted grating in a photonic crystal ring}

\author{Xiyuan Lu}\email{xnl9@umd.edu}
\affiliation{Microsystems and Nanotechnology Division, Physical Measurement Laboratory, National Institute of Standards and Technology, Gaithersburg, MD 20899, USA}
\affiliation{Joint Quantum Institute, NIST/University of Maryland, College Park, MD 20742, USA}

\author{Yi Sun}
\affiliation{Microsystems and Nanotechnology Division, Physical Measurement Laboratory, National Institute of Standards and Technology, Gaithersburg, MD 20899, USA}
\affiliation{Joint Quantum Institute, NIST/University of Maryland, College Park, MD 20742, USA}

\author{Ashish Chanana}
\affiliation{Microsystems and Nanotechnology Division, Physical Measurement Laboratory, National Institute of Standards and Technology, Gaithersburg, MD 20899, USA}

\author{Usman A. Javid}
\affiliation{Microsystems and Nanotechnology Division, Physical Measurement Laboratory, National Institute of Standards and Technology, Gaithersburg, MD 20899, USA}
\affiliation{Joint Quantum Institute, NIST/University of Maryland, College Park, MD 20742, USA}

\author{Marcelo Davanco}
\affiliation{Microsystems and Nanotechnology Division, Physical Measurement Laboratory, National Institute of Standards and Technology, Gaithersburg, MD 20899, USA}

\author{Kartik Srinivasan}\email{kartik.srinivasan@nist.gov}
\affiliation{Microsystems and Nanotechnology Division, Physical Measurement Laboratory, National Institute of Standards and Technology, Gaithersburg, MD 20899, USA}
\affiliation{Joint Quantum Institute, NIST/University of Maryland, College Park, MD 20742, USA}

\date{\today}

\begin{abstract}
    Frequency engineering of whispering-gallery resonances is essential in microcavity nonlinear optics. The key is to control the frequencies of the cavity modes involved in the underlying nonlinear optical process to satisfy its energy conservation criterion. Compared to the conventional method that tailors dispersion by the cross-sectional geometry, thereby impacting all cavity mode frequencies, grating-assisted microring cavities, often termed as photonic crystal microrings, provide more enabling capabilities through mode-selective frequency control. For example, a simple \ks{single period} grating added to a microring has been used for single-frequency engineering in \ks{Kerr} optical parametric oscillation (OPO) and frequency combs. Recently, this approach has been extended to multi-frequency engineering by using \ks{multi-period grating functions}, but at the cost of \ks{increasingly complex grating profiles that require challenging fabrication.} Here, we demonstrate a simple approach, \ks{which we term} as \ks{shifted grating multiple mode splitting} (SGMMS), \ks{where spatial displacement of a single period grating imprinted on the inner boundary of the microring creates a rotational asymmetry that frequency splits multiple adjacent cavity modes}. This approach is easy to implement \ks{and presents no additional fabrication challenges than an un-shifted grating, and} yet is very powerful in providing multi-frequency engineering functionality for nonlinear optics. We showcase an example where SGMMS enables OPO generation across a wide \ks{range of pump wavelengths} in a normal-dispersion device that otherwise would not support OPO.
\end{abstract}

\maketitle
\section{Introduction}
\noindent Nanophotonic platforms have enabled a wide variety of applications in optical systems engineering. One major advantage \xlt{of these platforms is their capability to achieve} complex functionality by photonic integrated circuits~\cite{Koch_PIC_1991, Nagarajian_JSTQE_2005}, which is often enabled by the precise control over the \yst{effective refractive index of the wavegudies}. Particularly in nonlinear optics, nanophotonic devices can enable phase matching of many nonlinear optical interactions that are challenging in bulk materials~\cite{Boyd2008} or fiber~\cite{Agrawal2007}. Generally, this is done by superimposing geometrical dispersion \yst{on} the nanophotonic waveguide to compensate for material dispersion~\cite{Agrawal2004}. In this regard, the whispering-gallery-mode (WGM) resonator has been the workhorse \ks{geometry} among all optical microcavities~\cite{Vahala_Nature_2003} for a wide variety of optical processes, including nonlinear optical interactions, optical sensing, and gain media for lasing~\cite{WGM-rev1,WGM-rev2,WGM-rev3}. These cavities provide tight confinement of light with high quality factors, resulting in an enhanced spatial and temporal enhancement for the intensity of the light field, which \yst{drives the efficiency of} nonlinear \yst{optical processes}~\cite{Boyd2008}; however, they impose strict constraints on frequency and phase-matching by requiring the interacting modes to have specific azimuthal mode orders and resonance frequencies to satisfy angular momentum and energy conservation, respectively. \ks{Such criteria have been satisfied, for example, in the application of chip-integrated WGM resonators to} a wide variety of $\chi^{(2)}$ and $\chi^{(3)}$ nonlinear processes such as optical parametric oscillation~\cite{Lu_Optica_2019,Bruch_Optica_2019,McKenna_NatCommun_2022,Lu_Optica_2020,Okawachi_NatCommun_2020,Reimer_NatCommun_2015} and frequency comb generation~\cite{gaeta_photonic-chip-based_2019,Chang_NatPhoton_2022}. In these cavities, dispersion can be \yst{engineered} by tuning the geometrical parameters, \yst{\textit{i.e.}}, the thickness, width, or the radius of the ring. This technique, however, can only control dispersion globally, \ks{with the resonance frequency of every mode of the ring being impacted}. 

\begin{figure*}[t!]
\centering\includegraphics[width=0.9\linewidth]{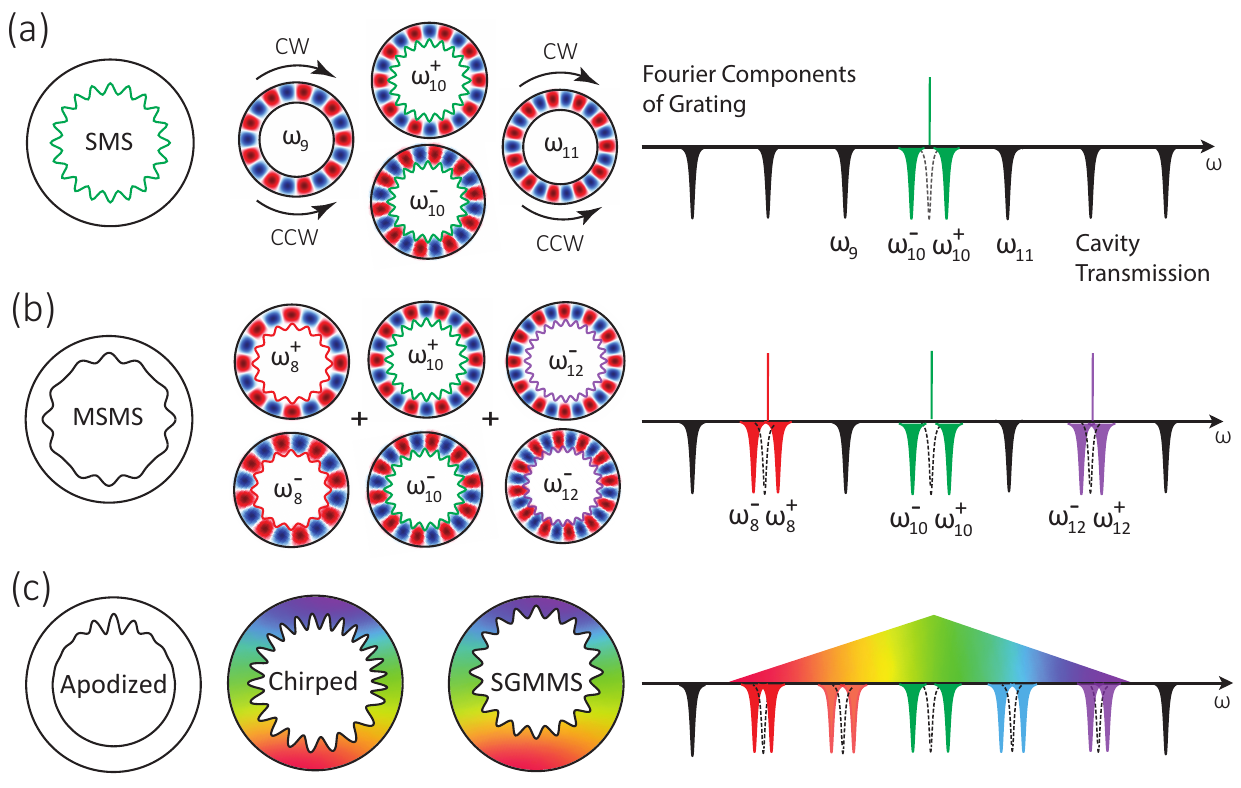}
\caption{ \textbf{Illustration of past techniques using grating-assisted microrings for frequency engineering and the current approach of shifted grating multiple mode splitting (SGMMS).} \textbf{(a)} In the single-frequency engineering case, a simple sinusoidal modulation of the inner boundary of the ring causes one mode, whose $m$-number matches with twice the modulation frequency, to \ks{have its resonance frequency split} into two. \yst{Other modes that are not matched can remain in their clockwise (CW) and counter-clockwise (CCW) traveling-wave propagation, and have no splitting in general.} This process is termed selective mode splitting~\cite{Lu_ApplPhysLett_2014}, as illustrated in the right panel of (a). \ks{Fourier analysis of the grating spatial frequency produces a delta function at the selected mode, and the microring cavity transmission displays that only the targeted mode is split.} \textbf{(b)} \yst{Selective mode splitting} can be extended to multi-frequency engineering by a simple sum of different spatial frequencies, termed as multiple selective mode splitting (MSMS)~\cite{Lu_PhotonRes_2020}, with the corresponding Fourier components and device transmission shown on the right. \textbf{(c)} An apodized or chirped grating, with more detailed Fourier analysis~\cite{Moille_DEMS_2023} or inverse design~\cite{Yang_NatCommun_2022}, can yield a continuous modulation of grating spatial frequencies, with the splitting amplitudes \ks{typically} varying in a continuous function versus mode frequencies. The shape of the grating profiles can be nontrivial to design and fabricate. We propose to circumvent these issues by using a simple single-frequency sinusoidal grating with an offset shifted from the ring center. The SGMMS technique can introduce controlled mode splitting in multiple modes by an intuitive design. We note that the \ks{displayed triangular function for the spectral profile of the mode splitting} is for illustration purpose only, and generally speaking, one can have various lineshapes depending on the specific grating parameters in use.}
\label{Fig1}
\end{figure*}

Another \yst{dispersion-engineering} technique uses sinusoidal modulation of the WGM resonator width to create back-scattering for a targeted mode \cite{Lu_ApplPhysLett_2014}. This technique, referred to here as selective mode splitting (SMS), can create sharp changes to dispersion at very specific wavelengths, where the resonator modes traveling clockwise \xlt{(CW)} and counterclockwise \xlt{(CCW)} have the correct momentum to be coupled by such coherent scattering \ks{induced by the grating modulation}. The coupling strength is manifested in the splitting of the mode frequencies, as illustrated in~Fig.~\ref{Fig1}(a). Here, the back-scattering can \ks{\yst{induce} a mode-selective change to} dispersion by causing the resonance to split into a doublet, shifting its frequency. This can be used to phase-match nonlinear interactions that otherwise would be mismatched in frequency~\cite{Lu_OptLett_2022,Black_Optica_2022,Stone_NatPhoton_2023,Yu_NatPhoton_2020}, \ks{where the width-modulated WGM resonator is also referred to as a photonic crystal ring}. The SMS approach can be extended to multiple modes by a simple sum of different modulation frequencies, each of which targets a specific mode, termed multiple selective mode splitting \cite{Lu_PhotonRes_2020}, as illustrated in~Fig.~\ref{Fig1}(b). This approach has recently been extended through Fourier synthesis to generate spatial domain modulation functions that target a large number of modes for broader-band applications~\cite{Moille_DEMS_2023}, as illustrated in~Fig.~\ref{Fig1}(c). Alternatively, one can also consider inverse design~\cite{Lucas_arXiv_2022, Yang_NatCommun_2022} to scatter specific momentum modes~\yst{by} a desired amount. These techniques~\cite{Moille_DEMS_2023, Lucas_arXiv_2022, Yang_NatCommun_2022}, though meticulous in their analysis, can produce spatial modulation functions that may be complex to implement, posing challenges in \ks{device design, fabrication, and the optical quality} of the resonances.

In this article, we introduce a simplified design approach to control dispersion of multiple modes using \yst{just a single} modulation frequency. We will refer to this \yst{method} as \yst{the} shifted grating multiple mode splitting (SGMMS). \yst{This approach is illustrated} in Fig.~\ref{Fig1}(c) labeled SGMMS, where we create an offset in the center of the spatial modulation of the ring width. In this technique, the modulation frequency targets a specific mode, as earlier, but the magnitude of the radial offset can control the \ks{number of modes that are \yst{split in frequency} (when the offset is zero, only the targeted mode is split)}. Therefore, a larger offset can induce mode-splitting of multiple modes without use of a complex multi-frequency modulation function. We demonstrate the \yst{capability} of this technique by simulating and experimentally demonstrating an optical parametric oscillator (OPO) in a silicon nitride ring resonator \ks{whose output colors are tuned by adjusting which of the SGMMS modes is pumped}. We show how several different pump modes from the same device, with a normal dispersion around \yst{the} pump, can simultaneously phase- and frequency-match the OPO interaction. \ks{In total, we find that the SGMMS technique offers a simplified design approach for dispersion control over multiple modes that is robust to fabrication constraints and maintains high optical quality factors. We anticipate that it can be applied to phase- and frequency-match \xl{any} cavity-enhanced parametric nonlinear optical processes.} 

\section{PRINCIPAL IDEA}
Grating technology, \ks{including not only uniform structures but also apodized and chirped geometries, has existed for decades, and specific implementations\yst{,} such as fiber Bragg gratings\yst{,} have} been widely used in sensing, filtering, and lasing~\cite{Raman_FBG_Book}. The topic discussed here, \ks{a grating inscribed} in a microring, \ks{has also been studied in many contexts. Some studies have focused on optical filtering, including for laser applications ~\cite{nozaki_ultralow_2003,arbabi_grating_2015,cheng_fully_2020}, while others have examined ejection into free-space orbital angular momentum (OAM) states~\cite{Lu_NatCommun_2023}, for example, in frequency comb applications~\cite{Chen_arXiv_2023, Liu_arXiv_2023}, where spectral control across multiple modes is also relevant~\cite{Moille_DEMS_2023,Lucas_arXiv_2022}. As the grating modulation strength increases, concepts from photonic crystals \yst{are increasingly involved}, and indeed, photonic crystal microrings (PhCRs) exhibiting slow light and defect-mode localization~\cite{Lu_NatPhoton_2022, Wang_PRL_2022} have been demonstrated.} Here, we focus \ks{our study of SGMMS for multi-frequency engineering in a regime where a perturbative single-frequency grating is imprinted upon a microring with a large and slow variation of its ring width.}

The context of SGMMS follows \ks{that of SMS~\cite{Lu_ApplPhysLett_2014}}. For instance, the ring width shown in~Fig.~\ref{Fig1}(a) is modulated with 20 periods per cycle targeting the optical mode with an azimuthal mode number of 10. This modulation results in coupling between the CW and CCW modes, breaking their frequency degeneracy. Consequently, two new standing-wave modes emerge at \ks{up-shifted and down-shifted} frequencies, as illustrated by two mode profiles labeled $\omega^+_{10}$ and $\omega^-_{10}$ in Fig.~\ref{Fig1}(a). The magnitude of the mode splitting $2\beta_{10}$, representing the frequency difference between the two modes, depends on the amplitude of the geometric modulation applied at the corresponding angular frequency ($A_{20}$). By precisely adjusting the modulation amplitude, the cavity frequencies can be accurately controlled. Additionally, when $m \ne 10$, where the modulation is out of phase with the optical mode, the optical mode remains unperturbed. For example, the $\omega_{9}$ and $\omega_{11}$ modes are unperturbed by SMS and effectively see only the average microring. This orthonormal property forms the basis of grating-assisted frequency engineering~\cite{Lu_ApplPhysLett_2014, Lu_PhotonRes_2020, Yang_NatCommun_2022, Moille_DEMS_2023}, and can be understood through perturbation theory of optical modes with shifted boundaries~\cite{Johnson_PRE_2002}, with:
\begin{eqnarray}
\beta_m = \frac{\omega_m}{2} \frac{\int{d S \cdot A \left[ (\epsilon_\text{c} - 1)|E_{\parallel}|^2 + (1 - 1/\epsilon_\text{c})|D_{\perp}|^2\right]}}{\int d V {\epsilon(|E_{\parallel}|^2 + |E_{\perp}|^2)} }, \quad \label{eq1}
\end{eqnarray}
where $E_{\parallel}$ ($D_{\parallel}$) and $E_{\perp}$ ($D_{\perp}$) are the electric field components (displacement field components) of the unperturbed optical mode \yst{polarizations} that are parallel ($\parallel$) and perpendicular ($\perp$) to the modulation boundary $d S$, respectively. \yst{And} $\epsilon$ represents the dielectric function of the material, including the silicon nitride core ($\epsilon=\epsilon_\text{c}$), silicon dioxide substrate, and air cladding ($\epsilon=1$).

\begin{figure*}[t!]
\centering\includegraphics[width=0.90\linewidth]{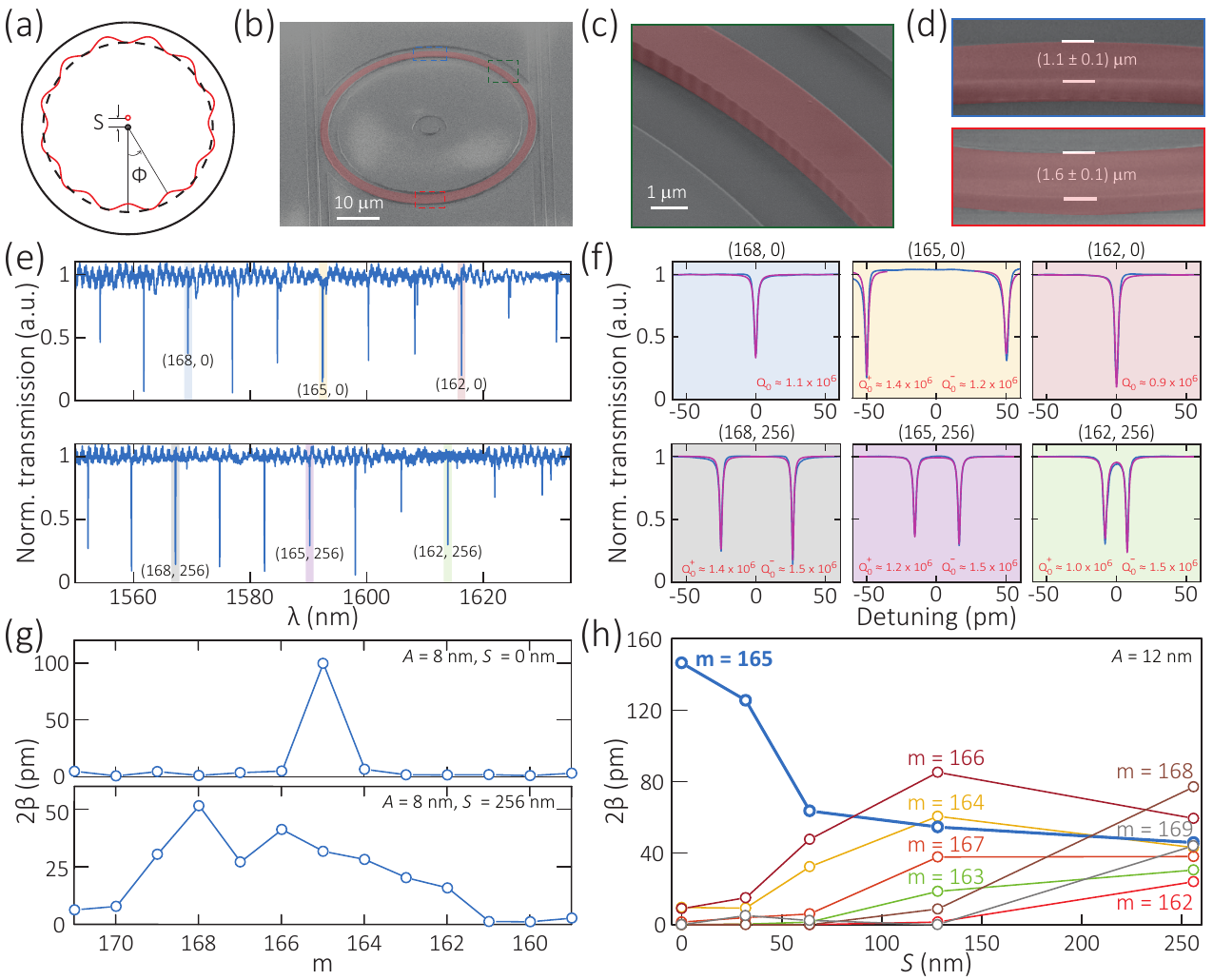}
\caption{\textbf{Experimental demonstration of SGMMS.} \textbf{(a)} Illustration of the SGMMS device with upward-shifted grating. \xl{The red and blue circles indicate the centers of the inside boundary (\yst{\textit{i.e.}}, the shifted grating) and the outside boundary of the microring, respectively.} \textbf{(b)} Scanning electron microscope (SEM) image of an SGMMS microring, false colored in red, with coupling waveguides on the left and right sides. In this article, only the right waveguide is used in experiments. The grating has a \xlt{nominal} shift $S$ = 256~nm. The \xlt{nominal} averaged ring width \yst{is} $RW$ = 1.5~$\upmu$m. \xlt{Zoom-in images of three parts of the microrings, highlighted by the dashed boxes, are shown in (c) and (d).} \textbf{(c)} \yst{An SEM image of the microring shows a sinusoidal modulation} with a \xlt{nominal} amplitude of $A$ = 12~nm. \textbf{(d)} The narrowest and widest $RW$s are measured to be (1.1 $\pm$ 0.1)~$\upmu$m and (1.6 $\pm$ 0.1)~$\upmu$m, respectively, \xlt{where the quoted uncertainty is the one standard deviation from multiple estimates from the SEM image}. \textbf{(e)} \xlt{Normalized} transmission spectra for the SMS device (top panel) with $A$ = 8~nm and $S$ = 0~nm and the SGMMS device (bottom panel) with the same $A$ but $S$ = 256~nm. The labels ($m$, $S$) represent the azimuthal mode number, and the shift in nanometers, respectively. \textbf{(f)} \yst{Scaled} transmission spectra \yst{from} (e) with \yst{annotations of the fitted $Q$ values}. In the top panel, the SMS device shows a mode splitting of \yst{approximately} 100~pm only at $m$ = 165 and negligible mode splittings at nearby modes. In the bottom panel, the SGMMS \ks{modulation introduces mode splittings across multiple modes}. The mode splittings for $m$ = 168, 165, and 162 are \xlt{approximately} 52~pm, 32~pm, and 16~pm, respectively. All these modes show high optical quality, with intrinsic $Q$ around 10$^6$. \textbf{(g)} A summary of the measured mode splittings (2$\beta$) across all the modes in (e). \textbf{(h)} The evolution of mode splittings from $S$ = 0~nm to $S$ = 256~nm, with a fixed $A$ = 12~nm. Above \ks{$S=50$~nm}, adjacent modes \yst{increasingly} start to show appreciable mode splittings. \ks{The uncertainties in mode splitting values in (g) and (h) are smaller than the data point size.}}
\label{Fig3}
\end{figure*}

When considering either transverse-electric-field-like (TE) modes or transverse-magnetic-field-like (TM) modes, the previous equation can be simplified even further. This simplification arises due to the dominance of the term involving either $D_{\perp}$ or $E_{\parallel}$ in the numerator's integral. \yst{Consider TE-like modes} as an example, where \ks{the modulation of the ring width for SGMMS is given by}:
\begin{eqnarray}
W(\phi) = W_\text{0} + A_{1} \text{cos}(\phi) + A_{2m} \text{cos}(2m\phi),  \quad \label{eq2}
\end{eqnarray} 
This SGMMS microring is illustrated by the third schematic in~Fig.~\ref{Fig1}(c). The azimuthal angle $\phi$ is illustrated in~Fig.~\ref{Fig3}(a).

When $A_{1}$ is perturbative, it has no effect \yst{on} $\beta_m$. The standing-wave mode with a larger frequency has a dominant displacement field $D(r,\phi,z)=D(r,z)\text{cos}(m \phi)$, and Equations~(\ref{eq1}-\ref{eq2}) lead to:
\begin{eqnarray*}
\beta_m = \frac{A_{2m} \omega_m}{2} \frac{ \int{d S (1 - 1/\epsilon_\text{c})|D(r,z)|^2 \text{cos}^2(m \phi) \text{cos}(2m \phi) }}{\int {d V \epsilon (|E_{z}|^2+|E_{\phi}|^2+|E_{r}|^2) \text{cos}^2(m \phi) } }. \quad \label{eq3}
\end{eqnarray*}
Here the azimuthal part can be integrated separately,
 \begin{eqnarray}
\beta_m = g_m A_{2m}/2, \quad \label{eq4}
\end{eqnarray}
where $g_m$ is defined as
\begin{eqnarray}
g_m \equiv \frac{\omega_m}{2} \frac{\int{d S (1 - 1/\epsilon_\text{c})|D(r,z)|^2}}{\int {d V \epsilon (|E_{z}|^2+|E_{\phi}|^2+|E_{r}|^2) }  }. \quad \label{eq5}
\end{eqnarray}

This linear dependence of mode splitting \yst{on} modulation amplitude only holds in the perturbative regime. A counter-intuitive observation is that the mode splitting can vanish to zero when the modulation amplitude increases to a critical value~\cite{Moille_DEMS_2023,Lucas_arXiv_2022}, \ks{and this effect can occur} regardless of the grating shape (sinusoidal, square, etc.) or \ks{photonic layer structure (cladding type, cladding symmetry, etc.)~\cite{Lu_arXiv_2023}. Such \yst{"}bandgap closing\yst{"} introduces additional complication when designing apodized gratings in microrings~\cite{Moille_DEMS_2023}. In this work, we restrict our analysis to the perturbative regime, where the mode splitting (bandgap) has a monotonic and linear dependence on modulation amplitude, to avoid this complication.}


In the SGMMS case, when the slow-variation term $A_1$ = $S/2$ is small \ks{(where $S$ is the shift relative to the center of the microring outside boundary)}, the splitting observed is expected to be close to the SMS case, with only a small amount of \yst{decreased mode splitting}. When $A_1$ is large, for example, the narrowest and widest parts of rings correspond to resonance frequencies over multiple free-spectral ranges, and other adjacent modes will \yst{also exhibit} mode splittings. Instead of an integration with $\phi$ from 0 to 2$\pi$, only a \yst{subset of this integration range} close to the targeted frequency would be relevant\yst{:}
\begin{eqnarray}
\beta_n \approx \frac{\Delta \phi_n}{2\pi} \frac{g_m A_{2m}}{2}, \quad \label{eq6}
\end{eqnarray}

\noindent \ks{where 2$m$ is the number of modulation periods within the ring circumference and $n$ is the mode under consideration.} We notice from Eq.~\ref{eq6} that the shifted-grating microring resembles a chirped-grating microring~\cite{Raman_FBG_Book}, as illustrated in~Fig.~\ref{Fig1}(c). The range of modes with splittings, that is, the range of $n$ with considerable $\beta$, is decided by the narrowest and widest ring widths, which is straightforward to determine in simulation. The value of $\Delta \phi$ is not as easy to calculate accurately, though intuitively easy to connect to~Fig.~\ref{Fig1}(c). We will measure the splittings experimentally in the next section.

\begin{figure*}[t!]
\centering\includegraphics[width=0.95\linewidth]{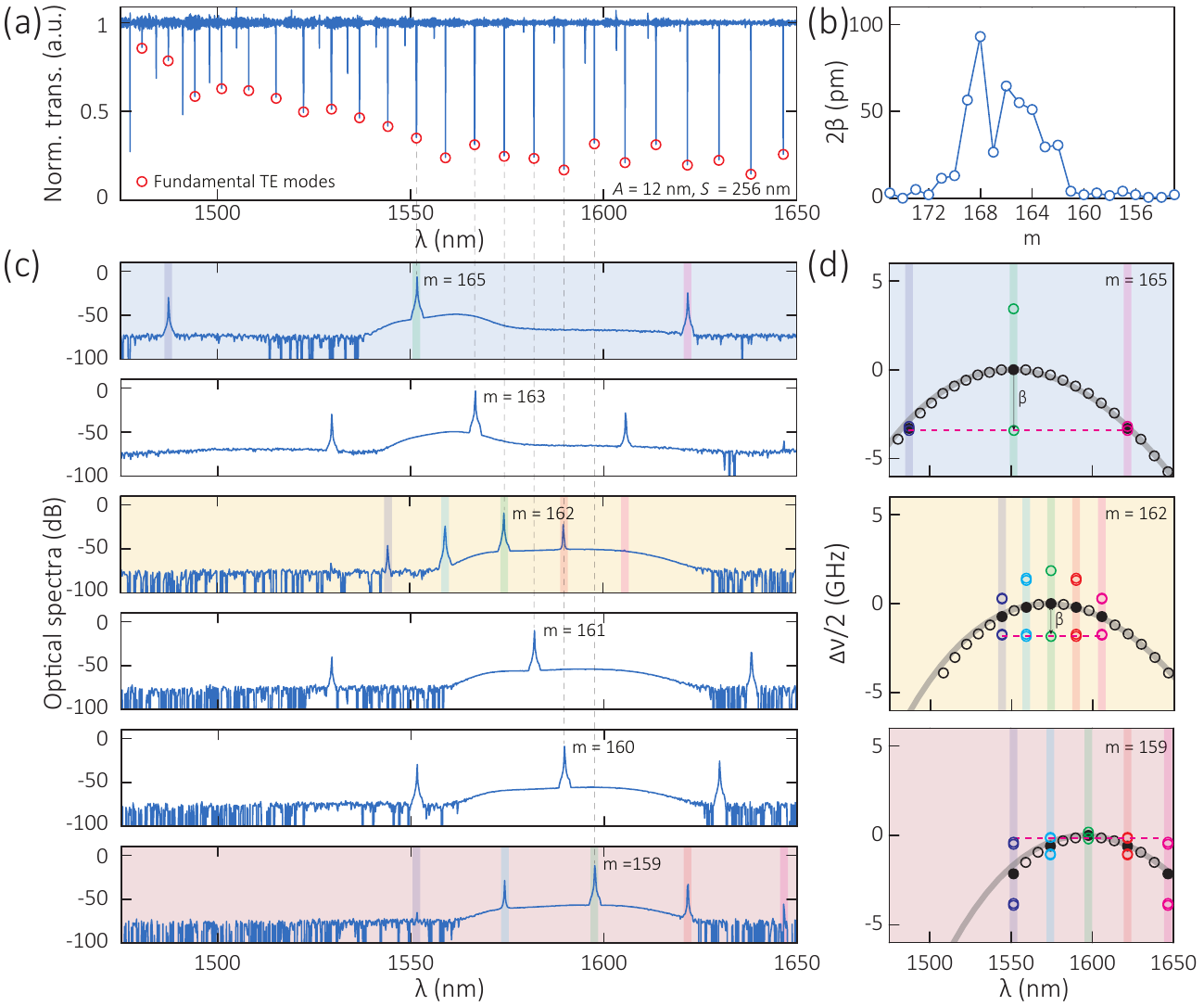}
\caption{\ks{\textbf{Application of SGMMS for pump-mode-selectable optical parametric oscillation (OPO).}} \textbf{(a)} Transmission spectrum of a fabricated SGMMS device with $A$ = 12~nm and $S$ = 256~nm. The resonance dips for the fundamental TE modes are circled in red. The target SMS mode is near 1552~nm with $m$ = 165. \textbf{(b)} A summary of the measured mode splitting values for the device in (a) from $m$ = 175 to $m$ = 153. With a shift $S$ = 256~nm, there are 10 modes (from $m$ = 171 to $m$ = 162) with mode splitting values larger than 10~pm. \textbf{(c)} The optical spectra from a series of OPO \ks{experiments, where the pump mode is varied between any of the six split modes at} $m$ = $\{$165, 163, 162, 161, 160, 159$\}$. \yst{On the y axis, 0 dB is referenced to 1 mW, \yst{\textit{i.e.}}, dBm.} \textbf{(d)} Frequency mismatch diagrams when pumping at $m$ = $\{$165, 162, 159$\}$. The grey curves are from finite-element-method simulations, and the circles are from experimental measurements. \yst{The uncertainties are smaller than symbol size} The mode splittings for relevant mode numbers are illustrated in colored circles, while their central \ks{(un-split)} frequencies are in solid black circles. \ks{Modes that are unrelated to the OPO process only have their central frequencies shown, in empty black circles. The frequency matching condition for OPO} is achieved at the red dashed lines. The colored columns highlighting involved modes in both (c) and (d) are for guidance of viewing.}
\label{Fig4}
\end{figure*}
\section{EXPERIMENTAL VERIFICATION}
We validate the SGMMS design using a high \yst{intrinsic} quality factor ($Q$) Si$_3$N$_4$/SiO$_2$ microring resonator platform \yst{from a} standard fabrication process~\cite{Lu_NatPhys_2019}. We implemented a grating modulation amplitude ($A$) on the inner side of the microring, with a shift ($S$) relative to the center of the outside boundary, as illustrated in~Fig.~\ref{Fig3}(a). The fabricated device\yst{, shown in Fig.~\ref{Fig3}(b-d),} has nominal outside ring radius, average ring width ($RW$), and Si$_3$N$_4$ thickness of 25~\textmu m, 1500~nm, and 600~nm, respectively. This device has \xl{nominal} $A$ = 12~nm and $S$ = 256~nm. This $S$ can be confirmed by half of the differences between the narrowest and the widest ring widths, as shown in~Fig.~\ref{Fig3}(d). 

We compare the SMS and SGMMS devices in~Fig.~\ref{Fig3}(e-g), and \ks{more closely examine} the details of multi-frequency engineering of SGMMS in~Fig.~\ref{Fig3}(h). We first examine the mode splitting behavior for the SMS device, that is, the control device for SGMMS \ks{(no shift)}. Here, we target the fundamental TE mode with $m$ = 165, which has a resonance around 1592~nm. This device has $A$= 8~nm and $S$ = 0~nm. The transmission spectrum of this device is shown in the top panel of~Fig.~\ref{Fig3}(e). \ks{The mode splittings of this device at all the modes are summarized in the top panel of~Fig.~\ref{Fig3}(g), and it shows the typical behavior of an SMS device.} There is only one notable mode splitting of \yst{approximately} 100~pm at the targeted mode $m$ = 165 and the mode splittings at nearby modes are all negligible with only less than 10~pm, in agreement with previous reports~\cite{Lu_PhotonRes_2020}.

We transform an SMS device into an SGMMS device by introducing the shift $S$. The bottom panel of~Fig.~\ref{Fig3}(e) shows the transmission spectrum of a fabricated SGMMS device with the same $A$ = 8~nm as the control SMS device and $S$ = 256~nm. As the shift is increased from 0~nm to 256~nm, the mode splittings are spread from the central mode at $m$ = 165 onto 8 modes with varying mode splittings of \yst{approximately} $\{$30, 52, 27, 41, 32, 28, 20, 16$\}$~pm at $m$ = $\{$169, 168, 167, 166, 165, 164, 163, 162$\}$, respectively, as shown in~Fig.~\ref{Fig3}(g). \xlt{One may also wonder} if there is any negative effect on the optical quality factors with such large shift $S$. Previous works~\cite{Lu_ApplPhysLett_2014,Lu_NatPhoton_2022} have shown that a grating with large modulations do not degrade the optical quality factors. In this work, we show that adding a large shift to this grating also maintains the $Q$s. The intrinsic $Q$s \yst{maintain values} over 1 million even with a shift $S$ = 256~nm as shown in~Fig.~\ref{Fig3}(f). We also investigate the \ks{behavior} of the mode splitting with various $S$ for another series of SGMMS devices with a fixed $A$ = 12~nm. \ks{In general, the number of modes with appreciable splitting (e.g., $>$10~pm) increases with $S$, while the maximum splitting amplitude for any mode occurs at $S$=0, where the entire effect of the grating is concentrated on one $m$. For example, the appreciable mode splittings are spread from the central mode to \ys{2} additional modes when $S$ = 64~nm, \ys{4} additional modes when $S$ = 128~nm, and eventually 7 additional modes (8 modes in total) when $S$ = 256~nm as shown in~Fig.~\ref{Fig3}(h).} Thus, the SGMMS approach introduces multiple frequency engineering capabilities with intuitive design and fabrication while maintaining high optical quality factor, all simply by adding a \ks{linear} shift to an SMS device. 

\section{OPO Application}
After validating the concept of SGMMS in the previous section, we now apply this technique \ks{directly to} frequency engineering problems in nonlinear nanophotonics applications. One important application is \ks{optical parametric oscillation (OPO), which converts an input laser from one color (\yst{\textit{i.e.}}, the pump) to two different output colors (\yst{\textit{i.e.}}, the signal and idler). OPO with the pump in either the normal or anomalous dispersion regime is possible; we focus on the former, where the signal-idler separation can be particularly broad and devoid of extraneous spectral tones~\cite{Lu_Optica_2019}. To realize such OPO, a close-to-zero but positive frequency mismatch $\Delta \nu = \nu_s+\nu_i-2\nu_p$ is required, where $\nu_{p,s,i}$ are the pump, signal, and idler modes respectively, as a result of the interplay between self- and cross-phase modulation of these three modes. The requisite frequency matching in a microring is typically realized through accurate cross-sectional geometric control, but \yst{this} comes with two main disadvantages: (1) the device is very sensitive to the geometric dispersion, which puts stringent requirements in design and fabrication; and (2) the close-to-zero dispersion requires the OPO to have a device layer thickness that is much larger than that typically offered within fabrication facilities for passive silicon nitride photonic circuits. Such thicker films typically require some level of additional fabrication (e.g., stress mitigation or damascene processing) to mitigate stress~\cite{Pfeiffer_JSTQE_2018}, which in turn complicates integration of the OPOs into more complex systems.}


The SMS approach has been applied to \ks{address the above challenges in previous works on OPOs and frequency combs \cite{Lu_OptLett_2022,Black_Optica_2022,Stone_NatPhoton_2023,Yu_NatPhoton_2020}.} However, this \ks{prior} approach enables OPO generation at the cost of \ks{constraining} the pump laser to be at a specific wavelength \ks{(mode)} defined by the SMS. Here we employ the SGMMS approach to extend the number of available modes for \ks{the pump} to enable OPO generation across \ks{the S+C+L bands (specifically from 1460~nm to 1625~nm).}  

\yst{To} take advantage of the full scanning range of our \ks{measurement} laser (from 1450~nm to 1650~nm), the SGMMS device characterized in~Fig.~\ref{Fig3} ($A$ = 12~nm and $S$ = 256~nm) is \yst{subsequently} etched from 600~nm to 500~nm\yst{,} to shift the SGMMS-central wavelength from 1592~nm to 1552~nm. Figure~\ref{Fig4}(a) shows the transmission spectrum of the etched device, \ks{with the corresponding mode splitting values} summarized in~Fig.~\ref{Fig4}(b). The SGMMS device shows a similar behavior as in~Fig.~\ref{Fig3}(g-h) and there are 10 modes (from $m$ = 171 to $m$ = 162) showing notable mode splitting values after etching. Though this device has normal dispersion across the telecom, \yst{all the red-shifted split modes}, can be pumped to generate OPOs~\cite{Lu_OptLett_2022}.~Figure~\ref{Fig4}(c) shows the experimental results of the OPO spectra as we pump \yst{six} of these split modes. When pumped at the red-shifted mode for $m$ = 165 (the top panel), we get a wide signal-idler span of 134~nm, in agreement with the frequency mismatch diagram in the top panel of~Fig.~\ref{Fig4}(d). This scheme is similar to \ks{prior SMS-OPO works}, \ks{where the mode splitting was implemented for only one mode, either the pump~\cite{Lu_OptLett_2022, Black_Optica_2022} or the idler~\cite{Stone_NatPhoton_2023}.}

Next, we leverage the \ks{multi-mode character of the} SGMMS technique, where we pump the \ks{split modes at different $m$ numbers and observe OPO for all \yst{modes}}.~Figure~\ref {Fig4}(c) shows the other 5 OPO spectra obtained when pumped at different modes accessible within our available amplifier range. For $m$ = 162 (the middle panel) and $m$ = 159 (the bottom panel), \ks{when pumped on the red-shifted resonances, two signal/idler pairs are observed and are observed at the red-shifted resonances for $m$ = 162 and at the blue-shifted resonances for $m$ = 159}. The data is in good agreement with the simulation and the measured \ks{frequency mismatch ($\Delta\nu/2$) as shown in the middle and bottom panels of~Fig.~\ref{Fig4}(d), where the frequency dashed red line indicates the mode combinations that satisfy the OPO frequency matching condition.} These two examples showcase the flexibility of the SGMMS approach in multi-frequency engineering. Such capability can advance the functionalities of \ks{other related nonlinear processes, including entangled photon-pair generation~\cite{Lu_NatPhys_2019,Graf_PRL_2022}, quantum frequency conversion via four-wave-mixing Bragg scattering~\cite{Li_NatPhoton_2016},} \xlpr{and frequency comb generation. In photon-pair generation~\cite{Yang_NatCommun_2022,Moille_DEMS_2023,Ulanov_arXiv_2023}, this method eases the task of creating mode splitting of the pump mode while maintaining high optical qualityfactor ~\cite{Lu_NatPhys_2019}, which leads to clockwise and counter-clockwise path entanglement of signal and idler photon pairs~\cite{Graf_PRL_2022}. In four-wave-mixing Bragg scattering, usage of the split mode for the pump laser can create the dispersion property required for quantum frequency conversion while also eliminating unwanted conversion channels~\cite{Li_NatPhoton_2016}. In frequency comb generation, our method simplifies the design of photonic crystal rings for both dispersion engineering~\cite{Yang_NatCommun_2022,Moille_DEMS_2023} and injection locking purposes~\cite{Ulanov_arXiv_2023}.}

\section{Discussion}
In summary, we demonstrate a nanophotonic frequency engineering approach \ks{that we term as} shifted grating multiple mode splitting (SGMMS), \ks{where controllable mode splittings across multiple resonator modes through a shifted, single-frequency grating}. SGMMS offers another level of simplicity in design and \yst{fabrication} over existing \ks{approaches}~\cite{Lu_PhotonRes_2020,Yang_NatCommun_2022,Moille_DEMS_2023}. We further \yst{apply} \ks{SGMMS-based} frequency engineering to \ks{optical parametric oscillation (OPO) across multiple pump modes in one microring device, with the \yst{generated} signal and idler output frequencies dependent on the pumped mode. Such results cannot be achieved using \yst{the previously demonstrated} single mode splitting devices~\cite{Lu_OptLett_2022,Black_Optica_2022} or through combinations of different (un-split) transverse mode families~\cite{Zhou_LPR_2022,Perez_NatCommun_2023}. While well-engineered OPO devices without the introduction of mode splittings can also support OPO on multiple pump modes, they have a more stringent tolerance of \yst{only} a few nanometers in \yst{the cross-sectional} geometry~\cite{Lu_Optica_2019,Lu_Optica_2020}, a result of the requirement to balance normal dispersion near the pump with higher-order dispersion for frequency matching. More generally, we anticipate that the introduction of a shift in the grating position shown in this work can be combined with \xl{previous} multiple-mode splitting techniques~\cite{Lu_PhotonRes_2020, Moille_DEMS_2023} and \xl{other dispersion engineering approaches to extend microcavity frequency engineering capabilities for nonlinear optical processes.}}


\medskip
\noindent \textbf{Funding.}
This work is supported by \ks{the DARPA LUMOS and NIST-on-a-chip programs}. X.L. acknowledges supports from Maryland Innovation Initiative.

\bibliographystyle{naturemag}
\bibliography{SGMMS}

\end{document}